# Quantifying electron irradiation effects in transmission electron microscopy

Toma Susi,[1] Jannik C. Meyer,[2] Jani Kotakoski[1]

1 University of Vienna, Faculty of Physics, A-1090 Vienna, Austria
2 University of Tübingen, Institute of Applied Physics, D-72076 Tübingen, Germany

**Abstract**: Important recent advances in transmission electron microscopy instrumentation and capabilities have made it indispensable for atomic-scale materials characterization. At the same time, the availability of two-dimensional materials has provided ideal samples where each atom or vacancy can be resolved. Recent studies have also revealed new possibilities for a different application of focused electron irradiation: the controlled manipulation of structures and even individual atoms. Evaluating the full range of future possibilities for this method requires a precise physical understanding of the interactions of electrons with energies as low as 15 keV now used in (scanning) transmission electron microscopy, becoming feasible due to advances both in experimental techniques and in theoretical models. We summarize the state of current knowledge of the underlying physical processes based on the latest results on two-dimensional materials, with a focus on the physical principles of the electron-matter interaction, rather than the material-specific irradiation-induced defects it causes.

## Introduction

The wavelength of optical photons restricts the resolution of light-based microscopy to scales far above interatomic distances. This fundamental limitation promoted the development of electron microscopy, which in the last two decades has broken through the last technical challenges to in principle resolve any interatomic spacing.[1,2] However, in addition to the shorter wavelength of electron waves that enables this resolution, as charged particles with mass, electrons also carry significant momentum and couple strongly to the electronic system of the material under study. Apart from the useful implications of such coupling for chemical identification through electron energy loss spectroscopy (EELS),[3,4] these interactions are in many cases a well-known detrimental side effect resulting in electron-beam damage through elastic collisions[5] with the nuclei as well as inelastic scattering that may result in ionization and radiolysis,[6-8] especially in organic samples.[9,10] Although such damage can be estimated from changes in the image contrast due to sample thinning,[11] or in the degradation of crystalline quality probed by electron diffraction,[10] quantification can only be fully precise when structural changes on the level of single atoms can be resolved.

Recent progress in sample preparation, especially of two-dimensional (2D) materials and their heterostructures,[12] have together with advances in the theoretical models of elastic collisions in some cases enabled their quantitative description,[13] which is more straightforward than in bulk crystals where damage cascades need to be considered.[14] However, inelastic scattering and its contribution to damage has been much harder to quantify.[10,15-18] Because of its increased probability at the lower electron energies now being used,[19-21] in many materials such damage is inevitable and hence new theoretical approaches are sorely needed. Beyond implications for electron-beam damage during imaging, these considerations have become increasingly important with the realization that electron irradiation can be used to sculpt materials,[22-27] to trigger phase transitions,[28-31] as well as to amorphize[32,33] or crystallize[34,35] structures. Very recently, it has even become possible to manipulate covalently bound atoms within materials,[36-41] allowing quantitative physical understanding to open new possibilities for materials science and nanotechnology.

To further these aims, our perspective summarizes the currently available knowledge of the mechanisms underlying electron irradiation effects. We start with a brief introduction into transmission electron microscopy and typical instrumentation, and then describe the physical principles of the electron-matter interaction[42] and the different excitations that can be induced.[43] The important role of atomic vibrations at low electron energies[13,44-46] as well as the often underappreciated interplay between the timescales of different excitations[47-49] with the time between successive electron impacts will be discussed for two-dimensional materials of widely varied dielectric properties.[49-51] We then survey the state of art in precision measurements and quantitative modeling, and close with our perspectives on the most important remaining challenges, missing experimental studies, and the realistically expected near-term future progress.

## Transmission electron microscopy

A thorough introduction into transmission electron microscopy (TEM) and scanning transmission electron microscopy (STEM) imaging is beyond our scope, but we refer the interested reader to recent textbooks.[52,53] Briefly, in TEM a parallel beam of electrons is transmitted through the sample and variations in the scattered intensity result in image contrast, whereas in STEM, a finely focused electron probe is rastered over the sample, with the recorded scattering intensity at each location forming the image. Here, we concentrate on the physical properties of the (mostly) relativistic electrons used as the microscopy probe, resolution in both space and time, and the characteristics of typical instrumentation in terms of beam current and current densities on the sample.

### *Electron irradiation*

Transmission electron microscopes use electron beams accelerated up to energies of typically 200–300 keV, though recent developments have focused on improving resolution at lower energies.[19-21] The relativistic velocity of 200 keV electrons is 0.69$c$ (where $c$ is the speed of light in vacuum), whereas for the lowest energies of 15 keV used in TEM it is 0.24$c$, with corresponding de Broglie wavelengths ranging from 0.027 to 0.10 Å, well below interatomic distances. The resolution of TEM is thus not limited by the physical characteristics of the radiation as in the case of diffraction-limited optical imaging, but rather by the technical limitations of the electron-optical lens systems. Another crucial difference is the momentum of the particles used for imaging: an electron accelerated to even 15 keV carries two orders of magnitude more momentum than a photon of the same energy, and can transfer multiple eVs of kinetic energy to an atomic nucleus when scattering to a high angle from its electrostatic potential.

### *Resolution in space and time*

Until the successful development of aberration correctors in the late 1990s and early 2000s to address imperfections in electron optics,[54,55] the only way to increase TEM resolution was to increase the electron energies, up to MeV to reach resolutions slightly worse than 1 Å.[56] However, with aberration correctors, typical instruments now allow imaging with resolutions between 1–2 Å at 60–80 keV, while the latest ones are able to reach 50 pm at 300 keV[57] or 80 pm at 80 keV with additional chromatic aberration correction.[58] As the current state of the art, such correction allows atomic resolution imaging for 2D materials to be retained down to 20 keV.[59] With post-processing techniques such as exit-wave reconstruction[60] or ptychography,[61] even higher resolution and more physical information can be obtained.

Time resolution, on the other hand, is mainly limited by the relatively low current densities of the electron beams, necessitating acquisition times of no less than fractions of a second to reach useful signal-to-noise ratios. Ultrafast techniques that aim to increase time resolution[62] are beyond our scope here, but we note that these are still rare and at significantly increased speeds their spatial resolution is limited (for example, a spatial resolution of 10 nm at a time resolution of ~0.1 fs, or 1 nm at ~0.1 ps).[63,64] Thus, as a general remark, atomic-level dynamics occur many orders of magnitude faster than the time between successive TEM images – indeed, even the typical STEM pixel dwell time of several μs is much too long to sample anything other than time-averaged metastable atomic configurations.

*Instrumental characteristics*
In STEMs, the scanning beam is focused to a spot of atomic dimensions, with a typical intensity profile full width at half maximum of ~1 Å.[65] A high probe current of 160 pA (50–120 pA is more typical[65,66]) corresponds to only one electron per ns, whereas the residence time of a 60 keV electron with a velocity of 0.45*c* is 15 ns in a 2 m long column. Thus the "beam" at the sample in reality consists of single electrons whose trajectories as an ensemble are focused by the lens system of the microscope. In TEM, on the other hand, the irradiated area is roughly equal to the imaged area, leading to dose rates per area that are 3–4 orders of magnitude lower even for a total beam current higher than 100 nA. The current can be focused to a spot of about 10 nm in diameter (although such a highly focused probe is no longer sufficiently parallel to be usable for TEM imaging), corresponding to a dose rate of $10^6$ $e^{-}Å^{-2}s^{-1}$, still at least an order of magnitude lower than for STEM.[67] An important difference between TEM and STEM is thus the local dose rate: the number of electrons that impinge on a unit area of the sample per unit time. This has implications especially for inelastic damage in non-metallic specimens due to the localization and the relaxation timescales of different excitations. Another important consideration for both instruments are the residual gases in the microscope vacuum (any molecules remaining in the instrument column at operating conditions) that may cause chemical etching[44] and even affect the relative stability of structures under electron irradiation.[68] However, etching is suppressed when the base pressure approaches ultra-high vacuum (≤$10^{-9}$ mbar).[69]

## Electron-matter interaction

As in any physical interaction, both momentum and energy need to be conserved when a probe electron interacts with a material, and since the electrons are moving at a fraction of the speed of light, relativistic effects must be accounted for. An elastic collision conserves kinetic energy and momentum, whereas an inelastic one conserves total energy and momentum, with part of the electron kinetic energy converted into excitations of the electronic (and vibrational) degrees of freedom in the material (Box 1). Even relatively low-energy electrons in the context of TEM would have more than enough kinetic energy to sputter atoms from any material, but momentum conservation in the (quasi-)elastic electron-nucleus scattering responsible for elastic knock-on damage severely limits the allowed energy transfers. However, already at electron energies below the expected threshold for displacing static atoms from the sample, the motion of a target atom may enable significantly greater energy transfers due to a summing of the initial momenta of the atom and the electron. For all practical purposes, relativistic electrons pass through a thin material so fast that they essentially interact with a frozen ionic and electronic system, with a timescale of zs ($10^{-21}$ s) estimated for elastic scattering from an atomic nucleus.[69]

Although sample morphology, including thickness, obviously plays a role in the details of the scattering process, we will use three of the most widely studied 2D materials as illustrative examples that span a wide range of dielectric properties. Hexagonal boron nitride (hBN) is an insulator a with band gap of close to 6 eV[70] and is heavily affected by ionisation;[71] monolayer molybdenum disulphide ($MoS_2$) is a 1.8 eV direct-gap semiconductor[72] that is thought to damage via a combination of ionisation and knock-on sputtering;[16] and finally graphene, a zero-gap semimetal[73] with ultrafast carrier mobility,[74] which is only affected by knock-on damage.[13]

**Box 1**: Elastic and inelastic damage.

In (quasi-)**elastic scattering**, an electron transfers momentum to an atomic nucleus, changing direction but losing little energy. Elastic scattering to small angles contributes to image contrast, whereas backscattering may cause knock-on damage.

In **inelastic scattering**, an electron couples to the electronic system of the material, losing some of its kinetic energy, but typically not deflecting significantly. In poorly conducting specimens, inelastic scattering contributes to damage via ionisation, radiolysis (direct breaking of chemical bonds) and/or charging.

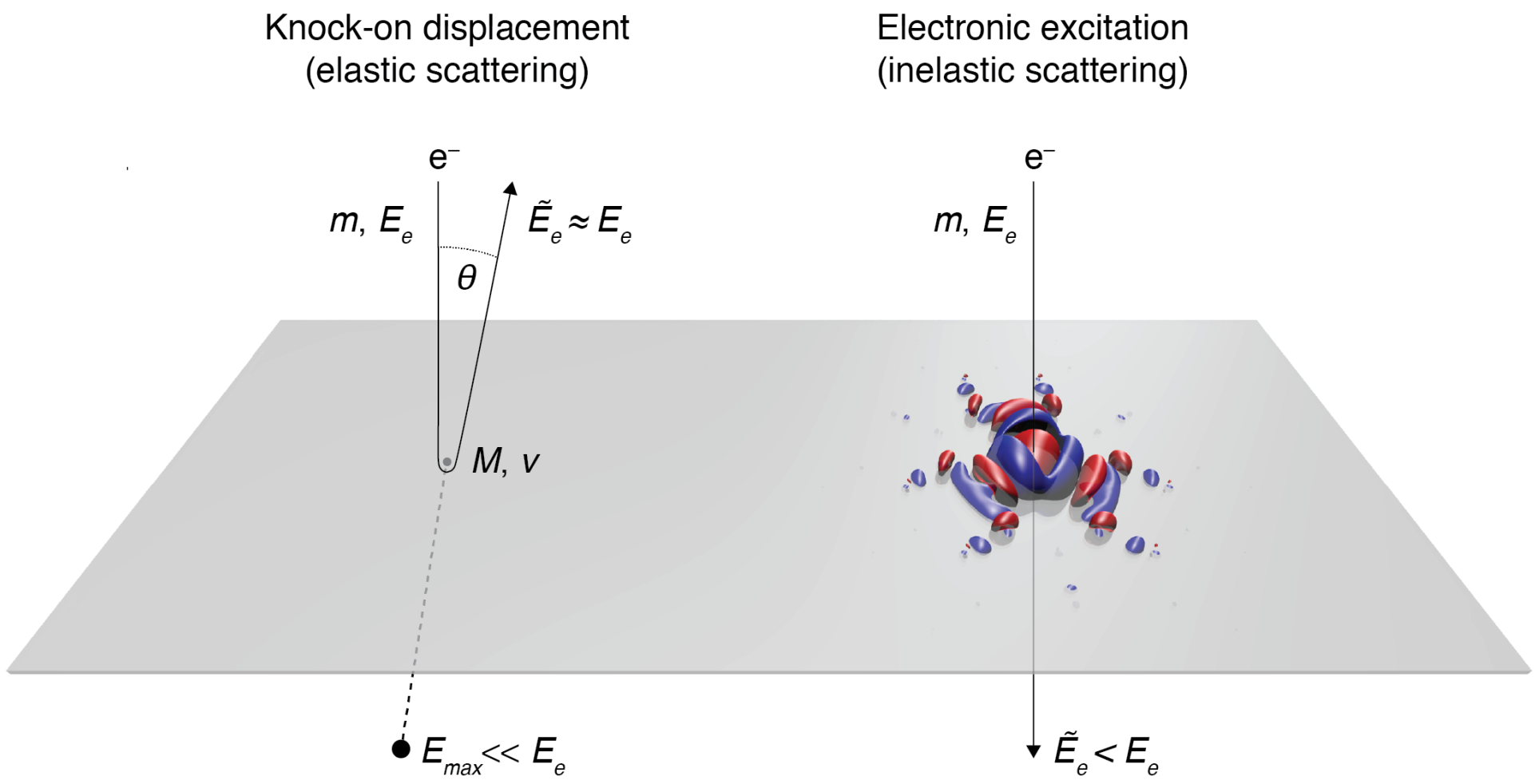

(For definition of the symbols, see Box 2.)

*Elastic scattering*

A few of the incident electrons scatter to large angles and transfer appreciable amounts of energy resulting in energy transfer $E$ allowed by relativistic energy-momentum conservation (Box 2). Considering only the static nucleus approximation and back-scattering, for our example materials with the commonly used electron energy of 60 keV, the electrons can only transfer a maximum of 11.61 eV to a $^{12}C$ nucleus, well below the graphene displacement threshold energy of 21.13 eV.[13] In hBN, 60 keV electrons can transfer 12.89 eV to B and 9.95 eV to N, with corresponding displacement threshold energies estimated with density functional theory molecular dynamics (DFT/MD) to be respectively 19.36 and 23.06 eV.[71] In $MoS_2$, these electrons can only transfer up to 4.35 eV to S and 1.45 eV to Mo, whereas the estimated thresholds are 7.1 and ~20 eV.[45,46] Although these values are well below the respective displacement thresholds of the materials, damage does occur in hBN and $MoS_2$ even below 60 keV,[75,76] highlighting the role of excitations in their damage mechanism.

**Box 2**: Elastic energy transfer.

Depending on the electron scattering angle $\theta$, relativistic energy-momentum conservation leads to elastic energy transfer

$$E(E_\mathrm{e},\theta,v)=\frac{2\left(E_\mathrm{e}\left(E_\mathrm{e}+2m_0c^2\right)+\sqrt{E_\mathrm{e}\left(E_\mathrm{e}+2m_0c^2\right)}Mvc\right)(1-\cos\theta)+(Mvc)^2}{2Mc^2} \quad \text{(I)}$$

for scattering of an electron with kinetic energy $E_\mathrm{e}$ and rest mass $m_0$ from an atomic nucleus with mass $M$ moving at velocity $v$ parallel to the beam ($c$ is the speed of light). Equation (I) has two illustrative limits: for a nucleus at rest ($v$ = 0), the well-known static formula is recovered:[79]

$$E(E_\mathrm{e},\theta,v=0)=\frac{E_\mathrm{e}\left(E_\mathrm{e}+2m_0c^2\right)}{Mc^2}(1-\cos\theta)=\frac{2E_\mathrm{e}\left(E_\mathrm{e}+2m_0c^2\right)}{Mc^2}\sin^2\frac{\theta}{2}, \quad \text{(II)}$$

whereas for backscattering ($\theta$ = 180°), the maximal energy transfer is given by[80]

$$E_\mathrm{max}\left(E_\mathrm{e},\theta=180^\circ,v\right)=\frac{\left(2\sqrt{E_\mathrm{e}\left(E_\mathrm{e}+2m_0c^2\right)}+Mvc\right)^2}{2Mc^2}. \quad \text{(III)}$$

To account for the motion of an atomic nucleus of mass $M$ moving at velocity $v$, the cross section for elastic scattering of a relativistic electron with kinetic energy $E_e$ can be evaluated from a power series expansion of the solution of the Dirac equation originally derived by Mott,[77] as further approximated by McKinley and Feshbach[42] accurate up to intermediate-$Z$ elements. Damage occurs when the transferred energy is greater than the displacement threshold energy $E_\mathrm{KO}$, resulting in a knock-on cross section (assuming an isotropic $E_\mathrm{KO}$):[78]

$$\sigma_{KO}(E_e,v)=4\pi\left(\frac{Ze^2}{4\pi\varepsilon_0 2\gamma m_0c^2\beta^2}\right)^2\left[\left(\frac{E_\mathrm{max}}{E_{KO}}-1\right)-\beta^2\ln\left(\frac{E_\mathrm{max}}{E_{KO}}\right)+\pi Z\alpha\beta\left\{2\left(\sqrt{\frac{E_\mathrm{max}}{E_{KO}}}-1\right)-\ln\left(\frac{E_\mathrm{max}}{E_{KO}}\right)\right\}\right], \quad (1)$$

where $E_\mathrm{max}(E_\mathrm{e},v)$ is given in Box 2, $Z$ is the atomic number, $e$ the elementary charge, the relativistic factor for the electron is $\beta(E_e)=\sqrt{1-(1+E_e/m_0c^2)^{-2}}$ with a corresponding Lorentz factor $\gamma=1/\sqrt{1-\beta^2}$, $m_0$ is the electron rest mass, and $\alpha$ is the fine structure constant. This can be numerically integrated over the atomic velocity distribution to obtain the total cross section, as explained below.[13,44]

*Importance of atomic vibrations*

Displacement threshold energies for atoms in the bulk of crystalline solids range from several to several tens of eV, whereas the kinetic energies of probe electrons are 3–4 orders of magnitude higher. Energy conservation thus clearly does not limit energy transfer in elastic scattering from the nuclear Coulomb potential. However, the electron is so light compared to atomic nuclei (~22 000 times lighter than carbon, for example) that the transferred energy is strictly limited by momentum conservation. In the static lattice approximation,[79,81] where it is assumed that the nucleus is at rest when the scattering occurs, energy transfer is limited by the maximal momentum change of the electron corresponding to backscattering.

However, the nuclei of atoms in any real material are not at rest due to thermal vibrations at finite temperatures as well as the quantum mechanical zero-point vibration.[44] Statistically, this motion can be described as a three-dimensional velocity distribution characterized by the Cartesian components of the mean square velocity. Although the average velocity in any direction is zero, each scattering electron essentially samples a single point of the distribution, whose tails correspond to rather high momenta. A large nuclear momentum in the direction of the electron beam allows total momentum to be conserved for much greater momentum transfers from the backscattering electron, thus enabling significantly more kinetic energy to be transferred.

The atomic vibrations have been described via the statistical distribution of out-of-plane velocities, characterized by its mean-square width (with the exception of an apparently equivalent description considering a single vibration period[46]). The width at each temperature $T$ has been estimated in three successively more precise ways: 1) assuming that the atomic velocities follow the ideal gas law,[45] 2) estimating the thermal population of phonon modes via a Debye model[44,76] (with Debye temperature $\theta_D$) or via a generic Bose-Einstein distribution[46], and finally 3) via an explicitly calculated out-of-plane phonon density of states (DOS) $g_z(\omega)$:[13,39,82]

Ideal gas: $$\overline{v_z{}^2}(T) = k_B T/M \, ; \tag{2}$$

Debye: $$\overline{v_z{}^2}(T) = \frac{3k_B}{8M}\theta_D + \frac{3k_B T}{8M}\left(\frac{T}{\theta_D}\right)^3 \int_0^{\frac{\theta_D}{T}} \frac{x^3}{\exp(x)-1} dx \, ; \tag{3}$$

Phonon DOS: $$\overline{v_z{}^2}(T) = \frac{\hbar}{2M}\int_0^{\omega_z} g_z(\omega)\left[\frac{1}{2} + \frac{1}{\exp\left(\frac{\hbar\omega}{kT}\right)-1}\right]\omega d\omega, \tag{4}$$

where $k_B$ is the Boltzmann constant, $\hbar$ the Planck constant, and $\omega_z$ the highest out-of-plane phonon frequency. Note that the above formulae for the ideal gas and Debye models include a factor of ⅓ to ensure correct partition of energy into the out-of-plane direction (missing from the original derivations in Refs. 44,45), which in the phonon model is accounted for by the normalization of the phonon DOS.

As an example, for graphene imaged at normal incidence, as shown in Figure 1, dynamics with calculated displacement threshold energies up to 3–4 eV higher than the maximum static energy transfer can be activated within experimental time scales due to out-of-plane vibrations at room temperature, as evidenced by experimental observations.[13,36,82,83] To actually calculate a cross section, the energy transfer as a function of the nuclear velocity needs to be integrated over the velocity distribution for those energies that exceed the displacement threshold.[13,46,82] It turns out that for most 2D materials imaged at electron energies of 80 keV or below, the effect of atomic vibrations must be included to quantify knock-on damage, or to even make correct qualitative predictions.

Thus far, the influence of atomic vibrations has only been included in the out-of-plane direction, *ie.* the nominal direction of the electron beam. While this may be sufficient to describe pure backscattering, the effects of tilt[82] and non-planar geometries[84] as well as specific dynamics such as the jumping of a pyridinic N atom laterally across a vacancy[85,86] will require that the theory is expanded to include in-plane components of the atomic vibrations and momentum transfer.

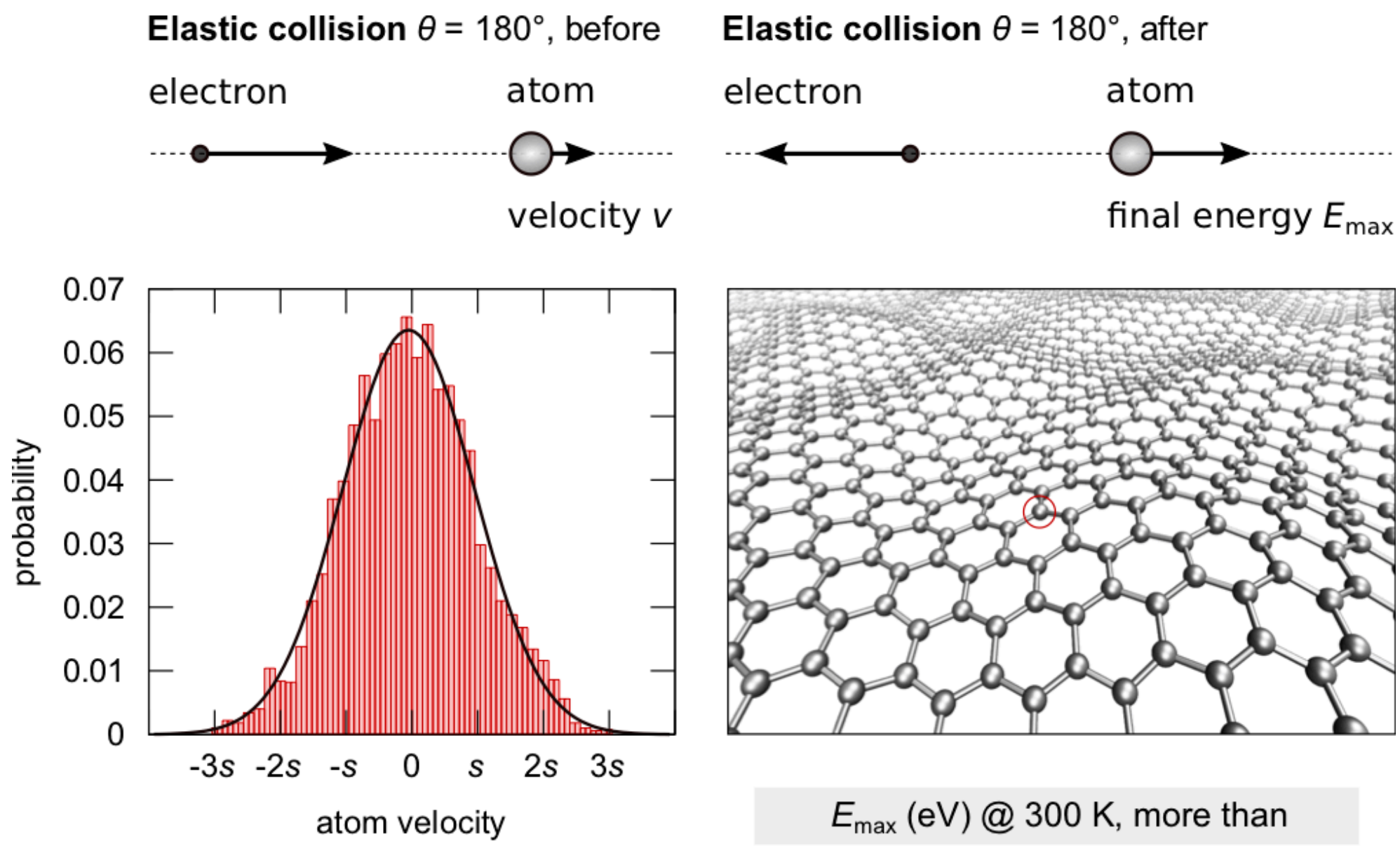

| | | $E_{max}$ (eV) @ 300 K, more than | | |
|---|---|---|---|---|
| | | 60 keV | 80 keV | 100 keV |
| 0.15% | $v > 3s$ | 14.66 | 19.30 | 24.03 |
| 2.5% | $v > 2s$ | 13.61 | 18.08 | 22.67 |
| 16% | $v > s$ | 12.59 | 16.91 | 21.35 |
| 50% | $v > 0$ | 11.62 | 15.77 | 20.08 |
| 84% | $v > -s$ | 10.68 | 14.68 | 18.84 |
| 97.5% | $v > -2s$ | 9.78 | 13.62 | 17.64 |
| 99.85% | $v > -3s$ | 8.92 | 12.61 | 16.48 |

**Figure 1.** Maximum energy transferred from an electron to a carbon atom in elastic backscattering ($\theta$ = 180°). Atoms in a lattice vibrate due to zero-point motion and the thermal excitation of phonons, which for two-dimensional materials leads to out-of-plane undulation of the structure, as seen in the snapshot of a molecular dynamics simulation. The distribution of out-of-plane velocities recorded for one atom once per fs during a finite temperature simulation over 5 ps follows a normal distribution whose standard deviation ($s$) is equal to the root mean square velocity of the atom. Depending on its velocity at the moment of electron scattering, the difference in the maximum transferred kinetic energy to a carbon atom at room temperature can vary by more than ±3 eV from the value corresponding to a static atom. The lower left panel illustrates the part of the velocity distribution that corresponds to the energy transfers given in the table of the lower right panel.

*Inelastic excitations*

As (typically relativistic) charged particles, electrons interact strongly with the electronic system of a material via the (retarded) Coulomb interaction. The bare electromagnetic field of an electron is evanescent,[43] unlike that of a photon, making the interaction highly localized and coupling particularly effectively to electrostatic and longitudinal excitations such as plasmons. Due to their thinness, however, most electrons pass through 2D samples without interaction, comprising the zero-loss peak seen in EELS.[3] Inelastic scattering for energy loss $E$ can be described via its double differential cross section, which depends on the energy-differential optical oscillator strength that is a function of energy loss but not scattering angle, and which decays monotonically with increasing energy proportional to $E^{-1}$ for large scattering angles $\theta$, and $E^{-3}$ for the very small $\theta$ that are typical for EELS.[3]

The most probable inelastic excitation is thus scattering from phonon modes in the material in the sub-eV regime, whose detection has recently become possible with the advent of capable electron monochromators.[4,87] With probability decreasing further with energy and depending on the material, inter- and intraband transitions and cathodoluminesence excitations occur in the few-eV range,[43] including valence ionization.[18] Next are plasmon excitations roughly in the 5–25 eV range,[88] which can still be quite intense even in atomically thin materials. Higher energy losses correspond to core level ionisation, which is useful for elemental identification and whose probability, overall decreasing with increasing energy loss, depends on the core ionisation cross section of each element and shell. Due to the quantum nature of the excitations, their localization is inversely proportional to their energy,[89] reaching atomic dimensions for higher core losses, several nm for optical excitations and plasmons, and several tens of nm for phonons.[4] However, very recent work with monochromated EELS has given new evidence for the localization of phonon effects down to the atomic scale.[90]

Inelastic excitations may contribute to damage through multiple routes, assuming that they do not relax back to the ground state faster than the relevant dynamical processes. Valence ionisation, either directly or via radiative or Auger recombination of a core hole, may weaken chemical bonds, activating knock-on damage at lower electron energies.[17] In some cases, an excited state may be converted into momentum and directly break bonds through radiolysis. This is a severe and dominant damage mechanism in molecules, organic materials, halides and silicates,[8] where the excitonic relaxation time is long enough and its energy large enough to cause atomic displacements. In insulators such as oxides irradiated at sufficiently high current densities, the locally induced electric fields may further directly contribute to phase transformations and damage.[8] In spatially confined structures insufficiently grounded by the sample support such as nanoclusters, the accumulation of charge may also result in damage if the resulting electrostatic repulsion overcomes bonding in the material in a ‘Coulomb explosion’,[91] whereas direct heating by the electron beam is negligible.[92]

Although precise theoretical descriptions for inelastic damage are still largely not available, a radiolytic atom displacement cross section can be estimated starting from the (relativistic) Bethe ionization cross section,[93,94] as given by Williams & Carter for each electron shell.[95] After converting to SI units, summing over all electrons by assuming the logarithmic term is equal for all shells, and using an identity for the kinetic energy, $T = (2\gamma m_0 c^2 \beta^2)/4$, we can get an approximation for the total radiolysis cross section:

$$\sigma_{\mathrm{rad}}\left(E_e\right) \approx 4\pi \frac{Ze^4}{\left(4\pi\varepsilon_0\right)^2 2\gamma m_0 c^2 E_r \beta^2}\left[\log\left(\frac{2\gamma m_0 c^2\beta^2}{E_r\left(1-\beta^2\right)}\right) - \beta^2\right]\zeta\,, \tag{5}$$

where $E_r$ represents a threshold energy that an electron of the material needs to receive to cause an atomic displacement. Its values for 2D materials are not known, but as a rough estimate, we may use the cohesive energy of each material: 7.46 eV/atom for graphene,[96] 5.11 eV/atom for $MoS_2$,[97] and 6.42 eV/atom for hBN.[98] The efficiency factor $\zeta$ (which now includes the contribution of shell-dependent factors) describes the proportion of ionisation events that contribute to bond breaking,[7,99] ranging from $10^{-4}$ for silicates to 0.1 for halides.[100] Its values or even the applicability of this model for 2D materials have currently not been determined due to the lack of quantitative experiments.

*Dynamical timescales*
Although both elastic and inelastic scattering occur in all materials, the dielectric nature of the sample determines whether inelastic excitations contribute to electron-beam damage. In good metals, including graphene, the relaxation of electronic excitations is extremely fast, and the material reaches its ground state between successive probe electrons. To illustrate the different timescales involved in inelastic excitations, we can consider either different excitations of the same material, or the same excitation in materials of varying dielectric properties.

Graphene provides the most widely studied example, with well-characterized excitation lifetimes: holes in the valence band are neutralized in $10^{-15}$ s, core holes in $10^{-14}$ s, whereas plasmons are damped within $10^{-13}$ s, and phonons in $10^{-12}$ s (Table 1).[47-49] Even for an intense STEM electron probe with a dose rate of $10^9$ $e^{-}s^{-1}$, only a single electron passes through the sample each $10^{-9}$ s, explaining why inelastic excitations do not play a role in electron-beam damage in this material. However, these assumptions may not hold locally at substitutional impurity sites.[86]

For a complementary perspective, we may consider the relaxation of excited electrons and holes in our three example materials. In graphene, the relaxation timescale is on the order of $10^{-15}$ s as mentioned above, whereas in semiconducting $MoS_2$, this time is close to $10^{-12}$ s, and in insulating hBN, as high as $10^{-9}$ s (Table 1).[49,51,101] Thus for hBN or $MoS_2$, or indeed practically any 2D material apart from graphene with its exceptionally high carrier mobility,[74] it can no longer be assumed that inelastic excitations do not contribute to the observed damage, though quantification of this remains a challenge.

**Table 1:** Order of magnitude lifetimes of inelastic excitations in 2D materials of differing dielectric properties (in seconds). The time that elapses between two consecutive electrons passing through the material is on the order of $10^{-9}$ s (for a high beam current of 160 pA).

| **Material** | Core hole | Valence hole | Plasmon | Phonon | Refs. |
|---|---|---|---|---|---|
| graphene | $10^{-14}$ | $10^{-15}$ | $10^{-13}$ | $10^{-12}$ | 47-49 |
| $MoS_2$ | $<10^{-15}$ | $10^{-12}$ | $10^{-10}$ | $10^{-9}$ | 101-104 |
| hBN | $10^{-15}$ | $10^{-9}$ | – | $10^{-12}$ | 51,105 |

**Quantification and modeling**
The ultimate precision for quantifying the effects of electron irradiation is to detect the displacement of single atoms, which is possible using atomic resolution TEM or STEM imaging of atomically thin specimens, especially 2D materials. The quantity that can be directly measured is the electron dose per sputtered atom – precisely speaking the expectation value of the underlying Poisson process, which is adequately described by the geometric mean dose.[36] This dose directly corresponds to a displacement cross section, from which a displacement threshold energy can then be extracted under the assumptions of McKinley and Feshbach[42] using Eq. 1. Despite the wide availability and interest in these materials, the quantification of damage is, however, only rarely reported,[13,15,16,44,67,76,86,106,107] and does require some care to correctly analyse even in nominally ideal circumstances.

An illustrative example are arguably the best understood experimental data, namely precision measurements of knock-on damage in pristine graphene. In a pioneering TEM study,[44] the rate of damage was estimated by counting the number of vacancies within the field of view after a known irradiation dose. However, especially at electron energies close to 100 keV, the damage was so fast that it resulted in multivacancy structures that no longer directly reflect the pristine displacement threshold (similar to later studies on the growth of holes in hBN[23,76]). Furthermore, some refilling of vacancies by diffusing carbon adatoms is expected,[108,109] which may have further affected the measurement. In a subsequent STEM study,[13] these issues were avoided by considering only tiny fields of view, counting the dose until the first atom was ejected, and scanning as fast as possible to not miss the healing of vacancies, which was indeed occasionally directly observed. Apart from the 100 keV datapoints, the TEM and STEM results agreed perfectly, demonstrating no dose-rate effect.[13]

Quantitative models of damage are still rare,[13,44-46,71,84,110-113] and thus far limited to knock-on damage apart from individual studies that have attempted to phenomenologically describe ionisation by including a localized charge.[71] Although computationally relatively cheap tight-binding or empirical potential calculations give reasonable results for pure carbon systems,[114] for systems with charge transfer such as hBN[71,84] or nitrogen-doped graphene,[83,115] these have been found to yield qualitatively incorrect results. This prompted a move to DFT/MD, which has become feasible with modern computation techniques and hardware. The choice of the exchange-correlation functional does seem to affect the predictions, with the local density approximation giving significantly higher energies than functionals based on the generalized gradient approximation. Whether weaker dispersion forces described by van der Waals functionals play a meaningful role remains uncertain,[76] though at least in the case of graphene, their inclusion results in the closest agreement with experiment.[13]

Regardless of the accuracy of the description of bond breaking, recent experiments at lower electron energies have demonstrated two facts. First, atoms are ejected at energies below the static limit, which can be accurately predicted for graphene when atomic vibrations are included. Second, damage of non-metallic materials cannot be described by purely elastic interactions.

Using the full quantum mechanical description of the phonons, a high predictive accuracy has been achieved for graphene[13] (correcting earlier predictions made via the Debye model[44]). However, significant discrepancies at graphene impurity sites between thusly simulated displacement threshold energies and those extracted from precision measurements have been noted and their origin remains unclear.[39,86] For $MoS_2$, the ideal gas and Bose-Einstein models appear to yield the same results (once thermal energy is correctly partitioned to only one Cartesian direction, Eq. 2)[45,46] – though it is clear that the damage is not purely knock-on as atoms are ejected already at 60 keV and below. Although quantitative data is just starting to emerge (Figure 2), the influence of imaging parameters such as dose rate will need to be assessed. For hBN, the best available theoretical description[76] uses a single Debye temperature,[116] which may be inadequate for a two-dimensional material.[44] Significant damage has been observed down to 30 keV, well below elastic displacement thresholds in the material, and its origin has been assigned to charging using variable temperature measurements.[76] Unfortunately, no reliable experimental estimate for the damage cross section of pristine hBN currently exists, since the available measurements[71,76] have only considered the growth of larger holes.

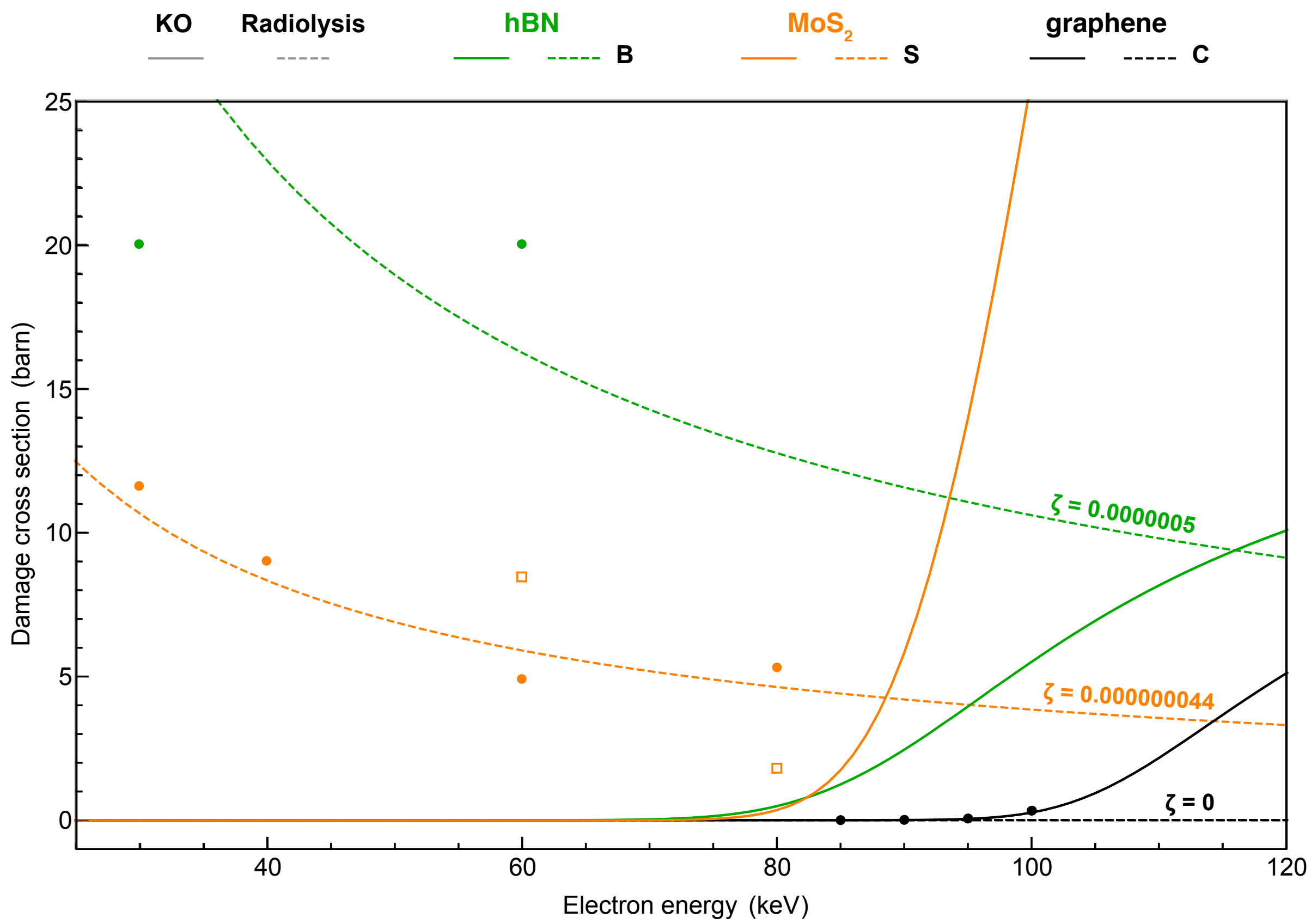

**Figure 2:** Damage cross sections from theory (lines) and atomically resolved measurements (points). The theoretical curves are by the authors using Eq. 1 for knock-on (KO) damage including the effect of vibrations via Eq. 2 for $MoS_2$, Eq. 3 for hBN, and Eq. 4 for graphene (solid lines), and Eq. 5 for radiolysis (dashed lines; efficiency parameters $\zeta$ estimated to match the data and overlaid next to the curves). Experimental points for graphene are from Ref. 13, for hBN from Ref. 75, and for $MoS_2$, from Ulm University (solid points; Tibor Lehnert, private communication), from the University of Chinese Academy of Sciences (open square at 60 keV; Wu Zhou, private communication), and from Ref. 45 (open square at 80 keV).

In principle, two-temperature models such as those used to describe electronic excitations in pulsed laser experiments[117] and swift heavy ion irradiation[118] might also be valid for electron irradiation effects in semiconductors and insulators. However, the applicability of such models for the present phenomena remains to our knowledge currently unexplored.

By combining 2D materials that suffer ionisation damage such as $MoS_2$ and hBN with highly conducting graphene in atomically thin van der Waals heterostructures, and controlling the orientation of the stack with respect to the electron beam direction, it has been possible to tease out differences in the damage rate and tentatively connect these with both elastic and inelastic damage channels. Pioneering experiments have been conducted with $MoS_2$/graphene[15,16] and $MoSe_2$/graphene heterostuctures,[17] revealing that not only do Se atoms sputter at electron energies far below their nominal displacement threshold, the mass of the chalcogen atom does not seem to affect the damage rate, clearly indicating a strong role of ionization. For hBN/graphene heterostructures,[76] the close vicinity of the conducting graphene monolayer appears to largely prevent damage to the hBN. However, more quantitative data at different electron energies is needed before these measurements can be reliably used in the development of theoretical models that include ionisation.

**Conclusions and perspectives**

We have summarized the current state of knowledge on the electron-matter interactions that are responsible for damage in transmission electron microscopy. Pure knock-on damage cross sections can in some cases now be accurately predicted completely from density functional theory molecular dynamics, with remaining questions related to the choice of the exchange correlation functional and the role of in-plane vibration components. Accurate and routine treatment of dynamics caused by elastic scattering into arbitrary angles remains a challenge, and new methods such as machine learning potentials may be needed to address large systems and long timescales.

However, electron irradiation damage is often either dominated or at least influenced by inelastic excitations, and their theoretical description remains difficult. It is unclear how well models proposed for damage in bulk oxides describe processes at the atomic scale, and how parameters such as radiolysis efficiencies could be predicted from first principles. Further, even in materials that can be described purely by knock-on damage in their pristine state, such as graphene, beam-induced processes at their impurity sites do not seem to allow for such simple treatment. Effects that may need to be incorporated include core and valence ionisation, an understanding of the lifetime of electronic excitations with respect to the electron dose rate, the possibility of multiple excitations by the same electron, and distinguishing direct bond breaking via charging or radiolysis from the weakening of bonds followed by elastic knock-on damage.

Precision measurements of irradiation damage at the level of single atoms in materials with varying dielectric properties at multiple and ever-lower electron energies are starting to emerge, and will provide much needed experimental guidance for theory. Two-dimensional materials offer in our view the best chance for providing the experimental data needed for further model development, which can then be adapted to materials more generally to provide a truly general and quantitative understanding of structural changes caused by electron irradiation. This will allow us to take full advantage of the emerging capabilities of Ångström-sized electron probes to sculpt and manipulate materials down to individual atoms, opening new possibilities for materials science and engineering.

**Acknowledgements:** We thank Tibor Lehnert, Ute Kaiser and Wu Zhou for providing $MoS_2$ data, and Arkady Krasheninnikov, Quentin Ramasse, Ovidiu Cretu, Anthony Yoshimura, Cong Su, Alexandru Chirita, Alexander Markevich and Gregor Leuthner for helpful discussions. T.S. was supported by the European Research Council (ERC) Grant 756277-ATMEN and the Austrian Science Fund (FWF) project P 28322-N36. J.C.M. was supported by the ERC Grant 336453-PICOMAT. J.K. was supported by the FWF projects I 3181 and P 31605, and the Wiener Wissenschafts-, Forschungs-, und Technologiefonds (WWTF) project MA14-009.

**Contributions:** T.S. conceived of the perspective and drafted the manuscript with contributions from J.C.M. and J.K. All authors revised the text.